\documentclass[twocolumn]{aastex63}

\usepackage{amsmath,amssymb,bbold,mathrsfs}
\usepackage{textcase}
\usepackage{subfigure}
\usepackage{xcolor}
\usepackage{soul}
\usepackage[nonumberlist,nosuper]{glossaries}
\setacronymstyle{long-short}

\definecolor{g}{RGB}{0,160,0}
\setstcolor{g}
\definecolor{b}{RGB}{0,0,250}
\setstcolor{b}
\definecolor{r}{RGB}{255,0,0}
\setstcolor{r}

\newcommand{\fref}[1]{Fig.~\ref{#1}}

\newcommand{\TAUE}[1]{${\rm log_{10}}(\tau_{500})=#1$}

\newcommand{\vlos}{$v_{\rm \parallel}$}
\newcommand{\blos}{$B_{\rm \parallel}$}

\newacronym{fwhm}{FWHM}{full width at half maximum}
\newacronym{los}{LOS}{line-of-sight}
\newacronym{clasp2.1}{CLASP2.1}{second flight of the Chromospheric LAyer 
SpectroPolarimeter}
\newacronym{tic}{HanleRT-TIC}{HanleRT Tenerife Inversion Code}
\newacronym{prd}{PRD}{partial frequency redistribution}
\newacronym{blos}{\blos}{magnetic field longitudinal component}
\newacronym{vlos}{\vlos}{velocity longitudinal component}
\newacronym{wfa}{WFA}{weak field approximation}

\received{}
\revised{}
\accepted{}

\shorttitle{}
\shortauthors{.}

\begin{document}

\title{Mapping the longitudinal magnetic field in the atmosphere of an active region plage \\
from the inversion of the near-ultraviolet CLASP2.1 spectropolarimetric data}

\author[0000-0001-5612-4457]{Hao Li}
\affil{Instituto de Astrof\'{\i}sica de Canarias, E-38205 La Laguna, Tenerife, Spain}
\affil{Departamento de Astrof\'\i sica, Universidad de La Laguna, E-38206 La Laguna, 
Tenerife, Spain}
\author[0000-0003-1465-5692]{Tanaus\'u\ del Pino Alem\'an}
\affil{Instituto de Astrof\'{\i}sica de Canarias, E-38205 La Laguna, Tenerife, Spain}
\affil{Departamento de Astrof\'\i sica, Universidad de La Laguna, E-38206 La Laguna, 
Tenerife, Spain}
\author[0000-0001-5131-4139]{Javier\ Trujillo Bueno}
\affil{Instituto de Astrof\'{\i}sica de Canarias, E-38205 La Laguna, Tenerife, Spain}
\affil{Departamento de Astrof\'\i sica, Universidad de La Laguna, E-38206 La Laguna,
 Tenerife, Spain}
\affil{Consejo Superior de Investigaciones Cient\'{\i}ficas, Spain}
\author[0000-0001-8830-0769]{Ryohko\ Ishikawa}
\affiliation{National Astronomical Observatory of Japan, 2-21-1 Osawa, Mitaka, Tokyo 
181-8588, Japan}
\author[0000-0001-9095-9685]{Ernest\ Alsina Ballester}
\affil{Instituto de Astrof\'{\i}sica de Canarias, E-38205 La Laguna, Tenerife, Spain}
\affil{Departamento de Astrof\'\i sica, Universidad de La Laguna, E-38206 La Laguna, 
Tenerife, Spain}
\author[0000-0002-9921-7757]{David E. McKenzie}
\affiliation{NASA Marshall Space Flight Center, Huntsville, AL 35812, USA}
\author[0000-0002-8775-0132]{Luca Belluzzi}
\affiliation{Istituto ricerche solari Aldo e Cele Daccò (IRSOL), Faculty of Informatics,
 Università della Svizzera italiana (USI), CH-6605 Locarno, Switzerland}
\affiliation{Leibniz-Institut f\"ur Sonnenphysik (KIS), D-79104 Freiburg i.~Br., Germany}
\affiliation{Euler Institute, Universit\`a della Svizzera italiana (USI), 
CH-6900 Lugano, Switzerland}
\author[0000-0003-3034-8406]{Donguk Song}
\affiliation{Korea Astronomy and Space Science Institute, 776 Daedeok-daero, 
Yuseong-gu, Daejeon 34055, Republic of Korea}
\affiliation{National Astronomical Observatory of Japan, 2-21-1 Osawa, Mitaka, 
Tokyo 181-8588, Japan}
\author[0000-0003-3765-1774]{Takenori J. Okamoto}
\affiliation{National Astronomical Observatory of Japan, 2-21-1 Osawa, Mitaka, 
Tokyo 181-8588, Japan}
\author[0000-0003-1057-7113]{Ken Kobayashi}
\affiliation{NASA Marshall Space Flight Center, Huntsville, AL 35812, USA}
\author[0000-0002-3770-009X]{Laurel A. Rachmeler}
\affiliation{National Oceanic and Atmospheric Administration, \\ 
National Centers for Environmental Information, Boulder, CO 80305, USA}
\author{Christian Bethge}
\affiliation{National Oceanic and Atmospheric Administration, \\ 
National Centers for Environmental Information, Boulder, CO 80305, USA}
\author[0000-0003-0972-7022]{Fr\'ed\'eric Auch\`ere}
\affiliation{Institut d'Astrophysique Spatiale, 91405 Orsay Cedex, France}

\begin{abstract}
We apply the HanleRT Tenerife Inversion Code to the spectro-polarimetric 
observations obtained by the Chromospheric LAyer SpectroPolarimeter. 
This suborbital space experiment measured 
the variation with wavelength of the four Stokes parameters in the near-ultraviolet 
spectral region of the \ion{Mg}{2} h \& k lines over a solar disk area containing 
part of an active region plage and the edge of a sunspot penumbra.
We infer the stratification of the temperature, the electron density, the 
line-of-sight velocity, 
the micro-turbulent velocity, and the longitudinal component of the magnetic 
field from the observed intensity and circular polarization profiles. The 
inferred model atmosphere shows larger temperature and electron density in the 
plage and the superpenumbra regions than in the quiet regions. 
The shape of the plage region in terms of its brightness is similar to 
the pattern of the inferred longitudinal component of the magnetic field in 
the chromosphere, as well as to that of the overlying moss observed by AIA 
in the 171 {\AA} band, which suggests a similar magnetic origin for the heating 
in both the plage and the moss region. Moreover, this heating is particularly 
significant in the regions with larger inferred magnetic flux.
In contrast, in the superpenumbra, the regions with larger electron density 
and temperature are usually found in between these regions with larger
magnetic flux, suggesting that the details of the 
heating mechanism in the chromosphere of the superpenumbra may be different to those 
in the plage, but with the magnetic field still playing a key role.
\end{abstract}


\section{Introduction} \label{intro}

The solar chromosphere is 
an extended region above the photosphere where the magnetic pressure
overcomes the gas pressure and dominates the dynamic and structure of the solar plasma
\citep[e.g., the review by][and references therein]{Carlsson2019ARA&A}. Unveiling and
understanding the magnetism of the solar chromosphere is key to discern
how the magnetic energy is transported to, and dissipated in, the chromosphere and
the transition region.

Our primary means to obtain 
empirical information on the solar magnetic field is via the inversion of the polarized 
spectra of atomic lines \citep[e.g., the reviews by][]{delToroIniesta2016LRSP,
Lagg2017SSRv,delaCruz2017SSRv}. Among the spectral lines forming in the upper solar
chromosphere, strong ultraviolet (UV) lines such as \ion{H}{1} Ly$\rm\alpha$ and 
the \ion{Mg}{2} h and k lines are considered promising 
for magnetic field diagnostics (\citealt{TrujilloBueno2011ApJ,TrujilloBueno2012ApJ,
Belluzzi2012ApJa,Belluzzi2012ApJb,Stepan2015ApJ,AlsinaBallester2016ApJ,
Tanausu2016ApJ,Tanausu2020ApJ,Trujillo2017SSRv}, and \citealt{Judge2022ApJ}).
However, acquiring the necessary UV spectro-polarimetric observations and inferring the
magnetic field from them is a rather challenging task
\citep[e.g., the review by][]{TrujilloBueno2022ARA&A}.

The sounding rocket experiments Chromospheric Lyman-Alpha 
SpectroPolarimeter \citep[CLASP;][]{Kobayashi2012ASPC,Kano2012SPIE} and 
Chromospheric LAyer SpectroPolarimeter 
\citep[CLASP2;][]{Narukage2016SPIE,Song2018SPIE} successfully observed 
the \ion{H}{1} Ly$\rm\alpha$ intensity and linear polarization in the quiet sun
\citep{Kano2017ApJ,Trujillo2018ApJ} and the \ion{Mg}{2} h and k 
full Stokes parameters in both
a quiet region close to the solar limb \citep{Rachmeler2022ApJ} and a plage region 
\citep{Ishikawa2021SA}, respectively (see also \citealt{Ishikawa2023ApJ}). 
By applying the \gls*{wfa},
\cite{Ishikawa2021SA} produced a map of the longitudinal component
of the magnetic field across the solar atmosphere, from the photosphere to the
upper chromosphere just below the transition region. 
Moreover, \cite{Li2023ApJ} applied the HanleRT Tenerife Inversion Code 
(hereafter, HanleRT-TIC)\footnote{The inversion 
code was simply dubbed TIC in previous 
publications \citep{Li2022ApJ,Li2023ApJ}. However,
since then, the inversion code has been completely integrated with the HanleRT
synthesis code \citep{Tanausu2016ApJ,Tanausu2020ApJ}, which used to be called
as a pre-compiled library during the Levenberg-Marquardt procedure.
HanleRT-TIC is publicly available at \url{https://gitlab.com/TdPA/hanlert-tic}.}
to invert the same data, obtaining a stratification of the temperature, electron 
density, line of sight velocity, micro-turbulent velocity, gas pressure, 
and the longitudinal component of the magnetic field at each location 
along the slit,
confirming and extending the results obtained 
by applying the WFA and further allowing for a rough
estimation of the energy that could be carried by Alfv\'en waves propagating across
the plage region chromosphere, which exceeded the expected radiative losses.

The success of these suborbital space experiments motivated the third
CLASP mission, the \gls*{clasp2.1}. Instead of sit-and-stare observations, 
the \gls*{clasp2.1} scanned a two dimensional 
field of view of an active region plage 
measuring the wavelength variation of the four Stokes parameters
in the same spectral region as the CLASP2 , i.e. from $\sim$279.30 to
$\sim$280.68~nm including the \ion{Mg}{2} h and k lines. 

Due to the formation height of  the \ion{Mg}{2} h and k lines,
the \gls*{clasp2.1} observation provides an opportunity to investigate the
heating processes in the upper chromosphere, close to the base of the
transition region to the corona. However, \gls*{prd} effects and
atomic level polarization induced by the scattering of anisotropic
radiation must be taken into account, even to model just their circular
polarization \citep{AlsinaBallester2016ApJ,Tanausu2016ApJ}.
These physical ingredients do not only add complexity to the the inversion of
the \ion{Mg}{2} h and k lines, but also significantly increase the computational
cost. In this work, we apply the HanleRT-TIC to invert these spectro-polarimetric
data and obtain the stratification of the temperature, electron density,
\gls*{los} velocity, micro-turbulent velocity, and the longitudinal
component of the magnetic field in the middle and upper chromosphere. 
To reduce the computing cost, we use a relative simple four-level Mg model atom,
which includes only the h and k lines. The model parameters resulting from the
inversion provide insights on the heating processes in the chromosphere and
the overlying corona.


\section{Observation} \label{sec2}

\begin{figure*}[htp]
  \center
  \includegraphics[width=1.\textwidth]{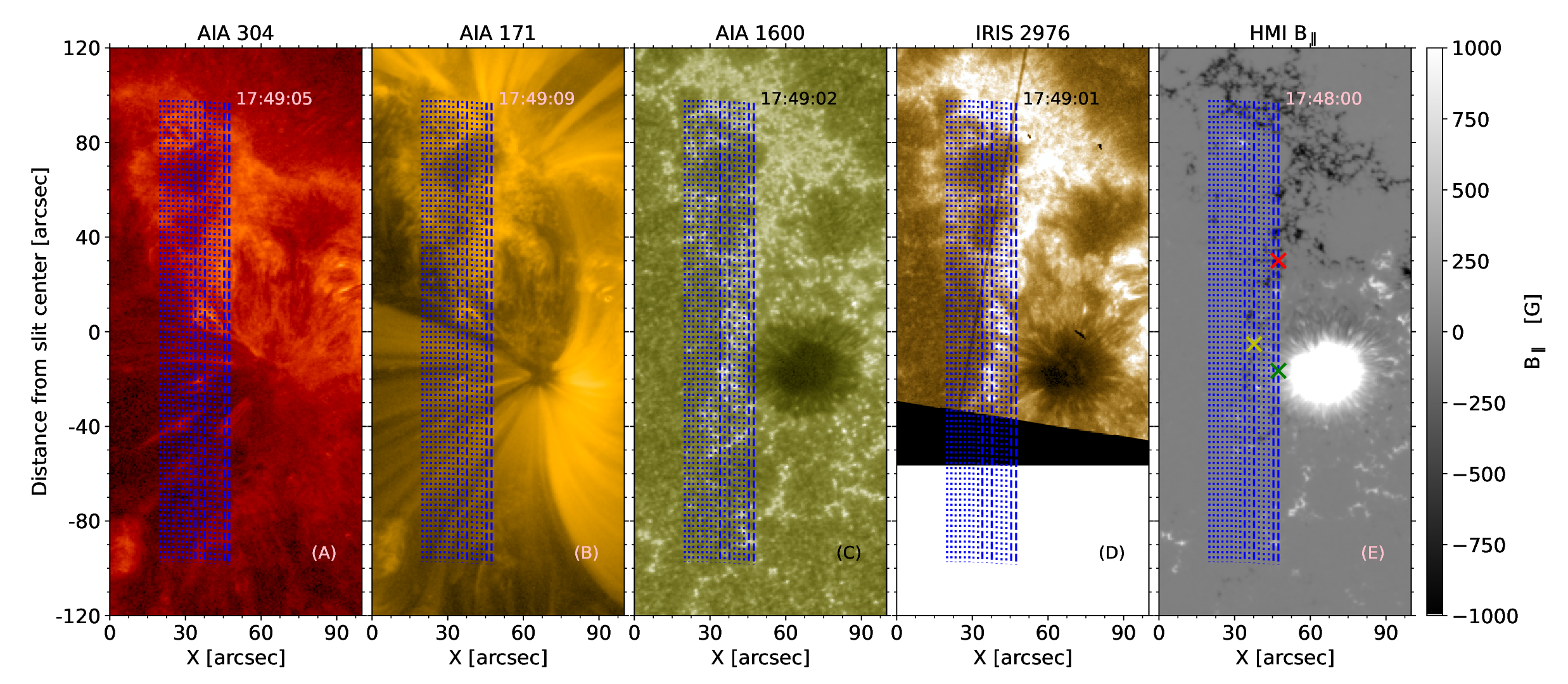}
  \caption{AIA images in the 304 (A), 171 (B), and 1600 {\AA} (C) bands.
  (D) IRIS slit-jaw image at 2796 {\AA}. (E) HMI magnetogram.
  Panels (A), (B), and (C) are shown in log scale. The blue vertical lines indicate 
  the 16 slit positions in the CLASP2.1 observation. The red, yellow, and green
  ``$\times$'' symbols in panel (E) indicate the position of the example profiles shown 
  in Figs.~\ref{fig4}, \ref{fig2}, and \ref{fig4a}--\ref{fig3a}.}
  \label{fig1}
\end{figure*}

The spectro-polarimetric data correspond to a spectrograph slit scan of 
NOAA AR 12882 performed by the \gls*{clasp2.1} suborbital experiment on 8 October 2021. 
The target active region was close to the disk
center, with a heliocentric angle $\theta$ such that $\rm\cos{\theta}\sim0.9$.
The scanning consists of 16 slit positions, acquired between
17:42:13 and 17:47:38~UT, with an exposure time of about 18~s which was
slightly different for each slit position. The spatial resolution along 
the slit direction is 0.53 arcsec/pixel. The slit raster step is about 1.8~arcsec on average.

The field of view covers a plage region and the edge of a sunspot penumbra,
with the same spectral range and resolution as the CLASP2, i.e. from $\sim$279.30 to
$\sim$280.68~nm with a spectral sampling of 49.9~m\AA/pixel, and a spectral point
spread function which can be approximated by a gaussian with a \gls*{fwhm} 
of about 110~m${\rm \AA}$ \citep{Song2018SPIE,Tsuzuki2020SPIE}. The full Stokes
spectro-polarimetric observation includes,
among other atomic lines, the \ion{Mg}{2} h and k lines, the two blended 
subordinate lines of the \ion{Mg}{2} at 279.88~nm, and three \ion{Mn}{1} 
resonance lines at 279.56, 279.91, and 280.19~nm. 
The wavelength is calibrated by assuming that the \ion{Mn}{1} lines are at rest. 
Therefore, the inferred longitudinal velocity values are relative to
the lower chromosphere, where the \ion{Mn}{1} lines are formed.
The polarimetric accuracy is abount $10^{-3}$ at the intensity peaks 
of the \ion{Mg}{2} h and k lines, which is slightly worse than that in the CLASP2 
observations due to the shorter integration time at each slit position, but still sufficient 
to apply the HanleRT-TIC in order to estimate the chromospheric magnetic field in 
the target region from the observed Stokes $I$ and $V$ profiles.

Observations of the target region obtained by the Atmospheric Imaging Assembly 
\citep[AIA;][]{Lemen2012SoPh} on board the Solar Dynamic Observatory 
\citep[SDO;][]{Pesnell2012SoPh} are shown in \fref{fig1}. The blue vertical 
lines in the figure indicate the 16 slit positions,
which cover a region of the active region plage and the edge of the sunspot
penumbra. 
A moss region \citep{Berger1999SoPh} located near 
the footpoints of some coronal loops
can be seen in the 171 {\AA} band approximately 
in the ranges 10 -- 60 ~arcsec in the X direction and
20 -- 100~arcsec in the Y direction.
The moss region also presents emission features in the 304 {\AA} band.
Coordinated observations of the same active region were acquired by 
the Interface Region Imaging Spectrograph satellite
\citep[IRIS;][]{DePontieu2014SoPh} and the Helioseismic 
and Magnetic Imager \citep[HMI;][]{Schou2012SoPh}.


\section{Inversion strategy} \label{sec3}

\begin{table}
  \centering
  \caption{Number of nodes in the temperature ($T$), micro-turbulent velocity 
  ($v_{\rm turb}$), vertical velocity ($v_z$), gas pressure ($P_{\rm g}$), and 
  vertical component of the magnetic field ($B_z$) for each inversion cycle. 
  Zero nodes indicate that the
  variable is fixed to the value resulting from the last cycle (or that it is set
  to zero if it is the first cycle). The single node in $P_{\rm g}$ indicates 
  that only its value at the top boundary is a free parameter and its stratification
  is obtained from the assumed hydrostatic equilibrium.}
  \begin{tabular}{c c c c c c} 
  \hline\hline
  cycle &  $T$   &   $v_{\rm turb}$   &    $v_{z}$  &  $ P_{\rm g}$   &  $ B_{z}$\\
  \hline
  1 & 4 & 3 & 3 &1 & 0 \\
  2 & 7 & 5 & 4 &1 & 0 \\
  3 & 0 & 0 & 0 &0 & 1 \\
  4 & 0 & 0 & 0 &0 & 4 \\
  \hline
  \end{tabular}  \label{tab1}
\end{table}

\begin{figure*}[htp]
\center
\includegraphics[width=0.9\textwidth]{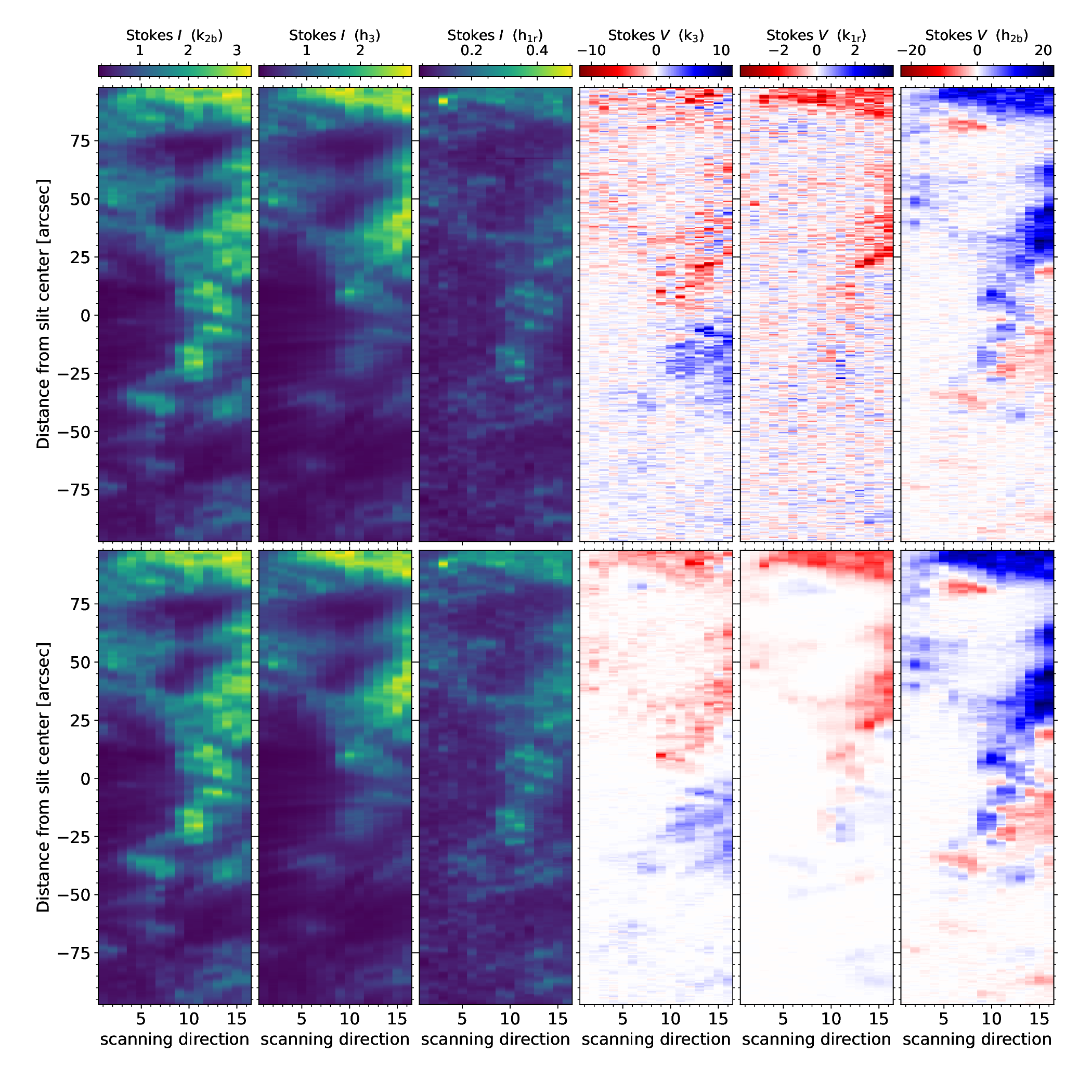}
\caption{Intensity at $\rm k_{2b}$ (first column), $\rm h_3$ (second column), and
$\rm h_{1r}$ (third column), and circular polarization at $\rm k_3$ (fourth column),
$\rm k_{1r}$ (fifth column), and $\rm h_{2b}$ (sixth column) for the observation
(top row) and the inversion fit (bottom row). 
The wavelengths corresponding to $\rm k_{2b}$, $\rm h_3$, $\rm h_{1r}$ $\rm k_3$, 
$\rm k_{1r}$, 
and $\rm h_{2b}$ are indicated with colored dashed lines in \fref{fig4}. 
The units for Stokes $I$ and $V$ are the same as in \fref{fig4}.}
\label{fig5}
\end{figure*}

In this paper we use the HanleRT-TIC,
a non-local thermodynamical equilibrium Stokes inversion code, 
which solves the radiative transfer problem 
assuming one-dimensional plane-parallel geometry. 
Hydrostatic equilibrium is assummed to compute the stratified gas pressure 
from that at the top boundary \citep{Mihalas1978}. The atomic and electron number densities
are calculated by solving the equation of state in local thermodynamic equilibrium 
with the method of \citet{Wittmann1974solphys}.
The code takes into account \gls*{prd} effects (in this paper we use the angle-averaged
formalism, see, e.g., \citealt{Mihalas1978,Leenaartsetal2012b,Belluzzi-TrujilloBueno2014}), 
$J$-state interference, and atomic level polarization 
in the incomplete Paschen-Back regime following
the formalism in \citet{Casini2014ApJ,Casini2017ApJ,Casini2017bApJ}. Note that
the atomic level polarization and the radiation field anisotropy significantly 
impact the circular polarization of the outer lobes 
of the \ion{Mg}{2} h and k line profiles
\citep{AlsinaBallester2016ApJ,Tanausu2016ApJ}.

\begin{figure*}[htp]
  \center
  \includegraphics[width=0.9\textwidth]{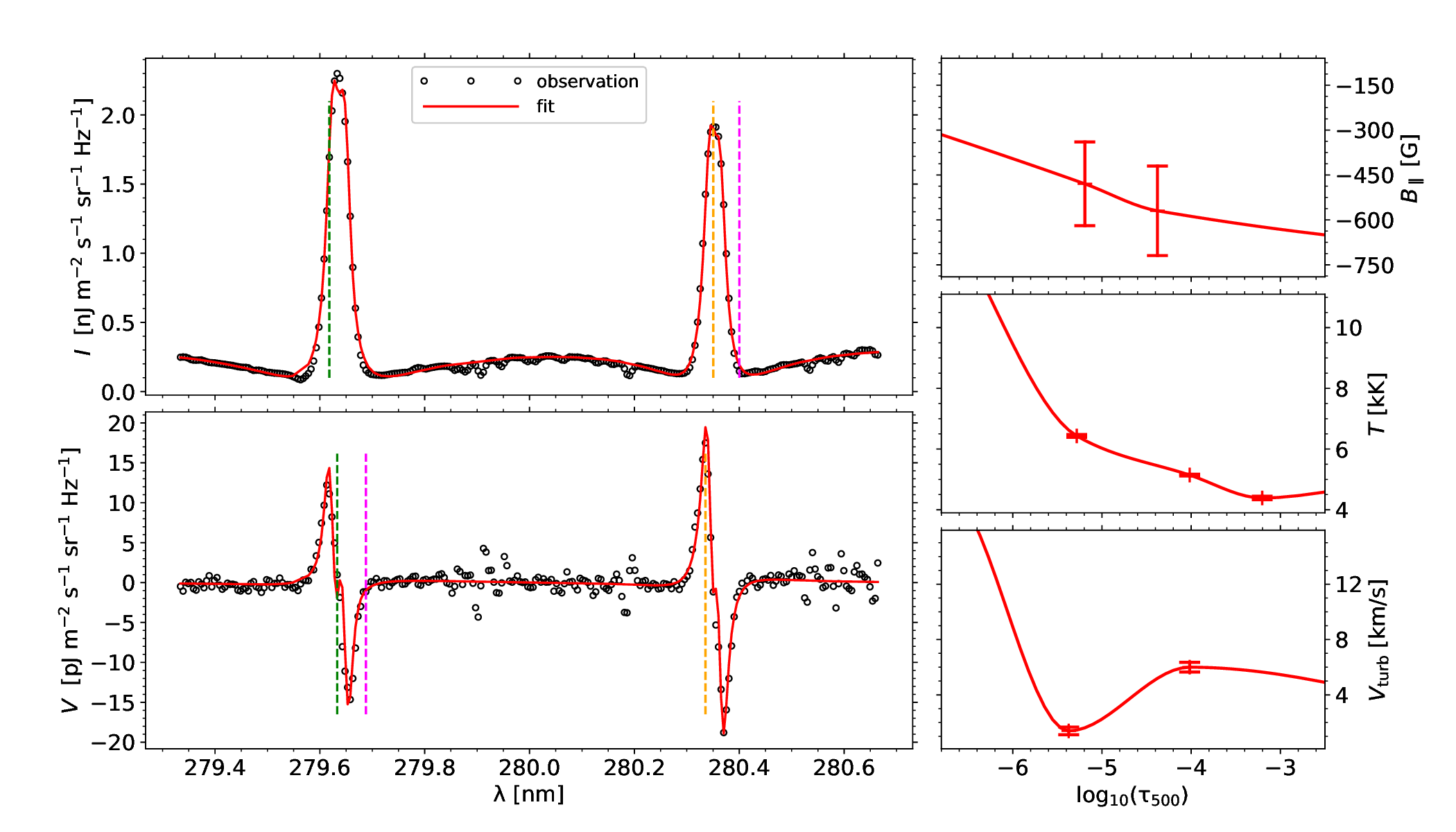}
  \caption{ Observed (open circles) and fitted (red solid curve) 
  intensity (top left panel) and circular polarization (bottom left panel) profiles for 
  a pixel in the plage (red ``$\times$'' symbol
  in Panel (E) of \fref{fig1}). Inferred $B_\parallel$, $T$, and $v_{\rm turb}$
  are shown in the right panels, from top to bottom, respectively.
  The ${\rm k_{1b}}$ minimum (at around 279.57 nm) is blended with a \ion{Mn}{1} line,
  and those frequency nodes are not present in the inversion (thus resulting in a
  straight line in the plot). The green, orange, and purple dashed lines in the top
  left panel indicate the wavelength location of ${\rm k_{2b}}$, ${\rm h_3}$,
  and ${\rm h_{1r}}$, respectively. The green, orange, and purple dashed lines in the
  bottom left panel indicate the wavelength location of ${\rm k_3}$, ${\rm h_{2b}}$,
  and ${\rm k_{1r}}$, respectively.}
  \label{fig4}
\end{figure*}

\begin{figure*}[htp]
  \center
  \includegraphics[width=0.9\textwidth]{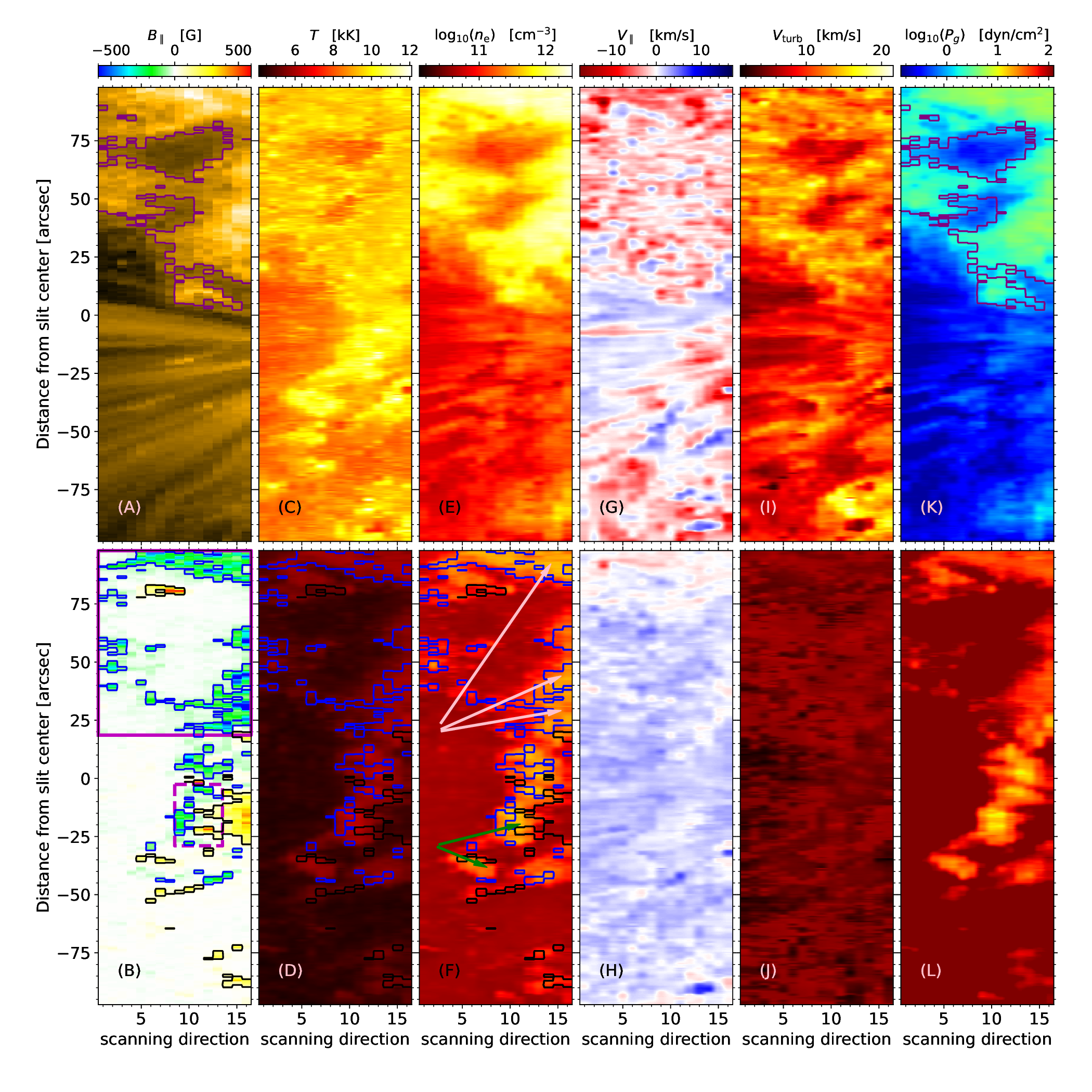}
  \caption{Top left panel: AIA image in the 171~\AA\ band.
  Bottom left panel: HMI magnetogram (corresponding color bar at the top of the 
  top-left panel). From the second to the sixth columns: Inferred 
  temperature, 
  electron density, longitudinal component of the velocity, micro-turbulent velocity, 
  and gas pressure at \TAUE{-6.0} (top panels) and \TAUE{-4.5}, (bottom panels). 
  Purple contours in panels (A) and (K) are drawn according to the intensity 
  of the 171 {\AA} AIA band. 
  Black and blue contours in panels (B), (D), and (F) correspond to the HMI
  magnetic field values of 70~G and $-70$~G, respectively. In panel (B),
  the squares drawn with solid and dashed purple lines delimit the pixels included in
  the scatterplots in \fref{fig7} (see text). The white and green arrows highlight
  regions of the field of view discussed in the text.}
  \label{fig6}
\end{figure*}

\begin{figure*}[htp]
  \center
  \includegraphics[width=0.8\textwidth]{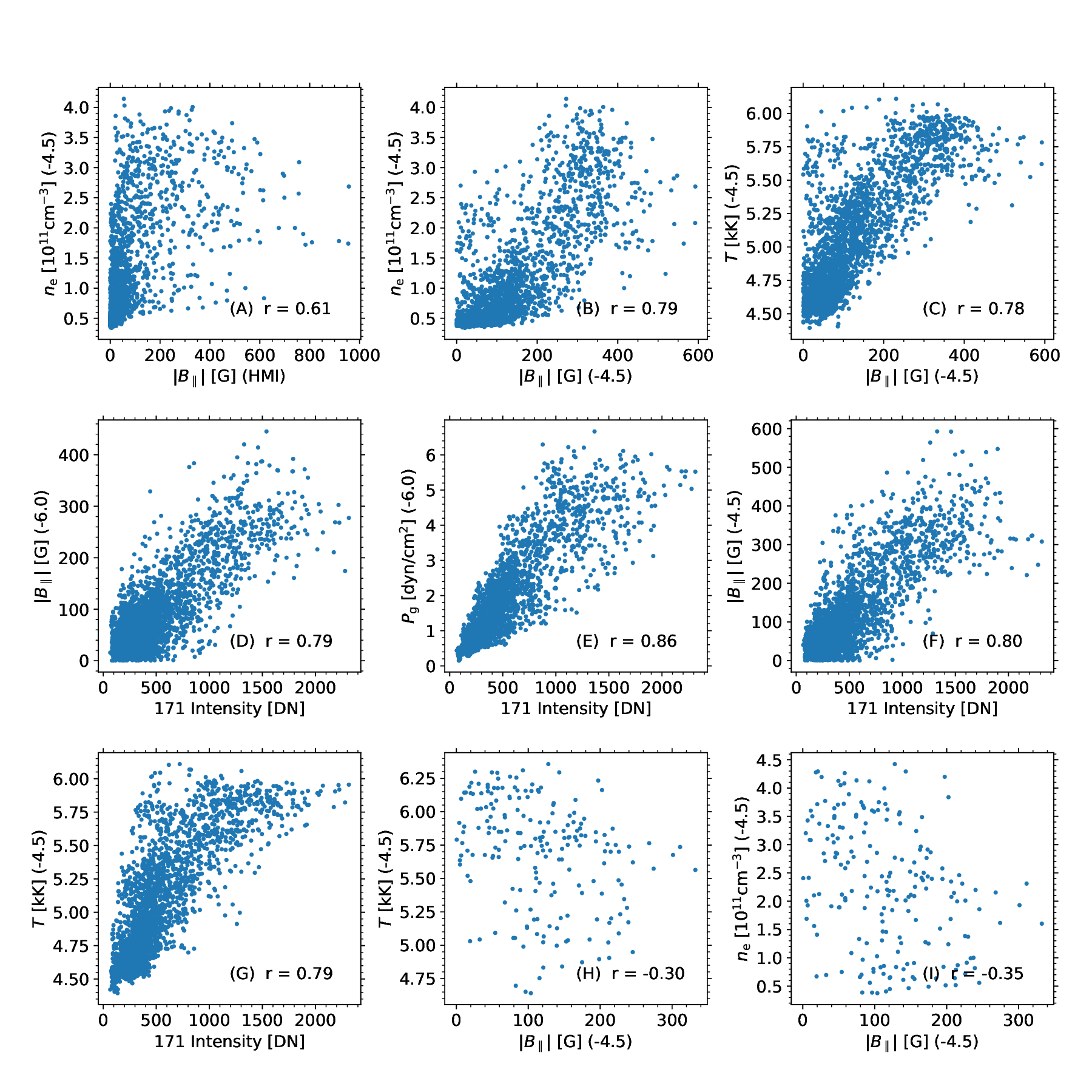}
  \caption{Scatterplots for selected parameter pairs. In the labels, the number
  between parenthesis indicate the $\log_{10}(\tau_{500})$ where the value of the 
  variable is taken from the inversion result, ``(HMI)'' indicates that the value 
  is taken from the HMI magnetogram, and the label ``171 Intensity [DN]'' indicates 
  the intensity in the AIA 171~\AA\ band. The points in panels 
  (A) -- (G) correspond to the region delimited by solid purple lines in the 
  bottom-left panel of \fref{fig6}, while the points in panels (H) and (I) 
  correspond to the region delimited by the dashed purple lines in the same 
  panel of \fref{fig6}. The Pearson correlation
  coefficient is shown in each panel.}
  \label{fig7}
\end{figure*}

\begin{figure*}[htp]
  \center
  \includegraphics[width=1.0\textwidth]{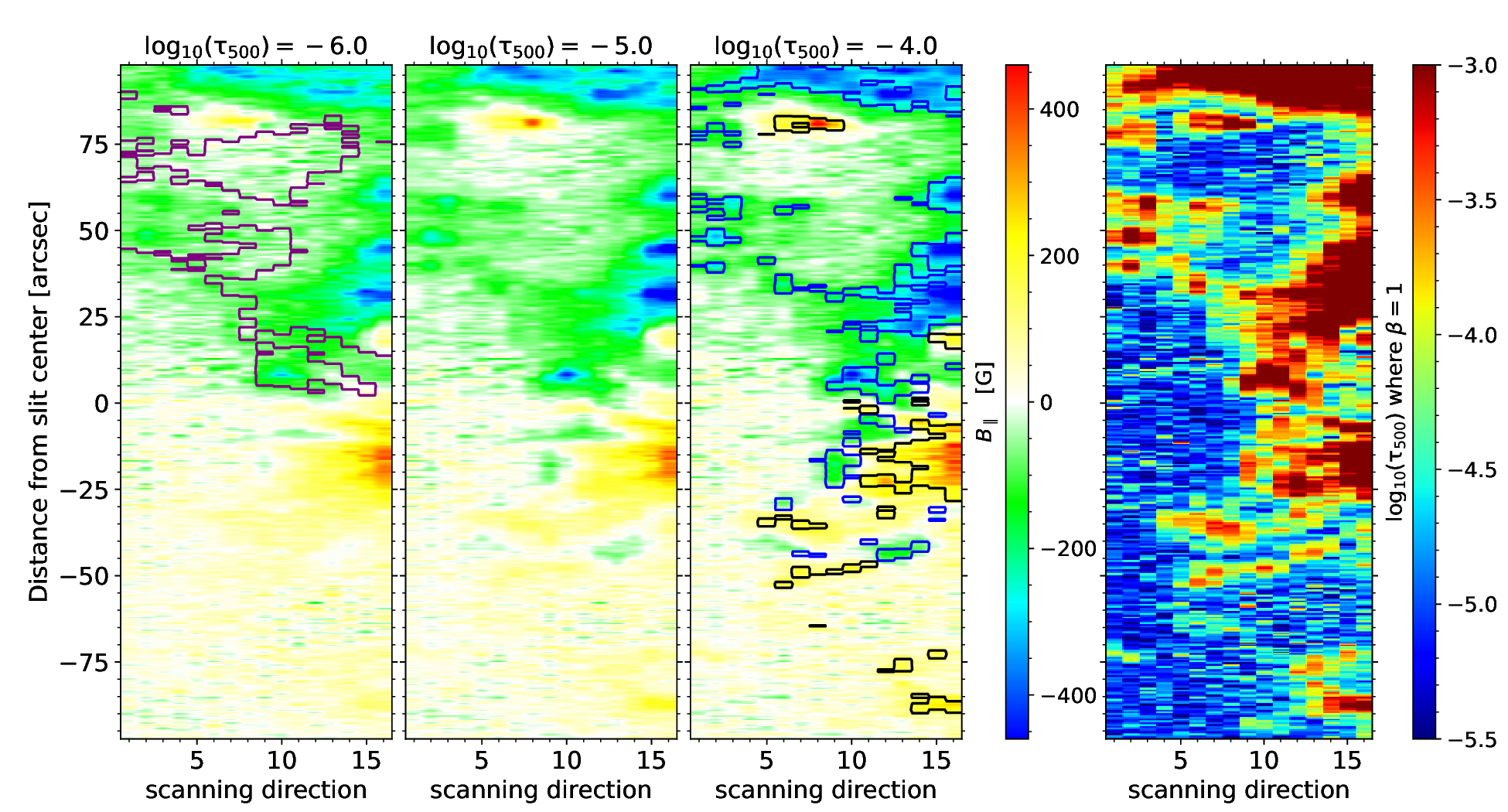}
  \caption{Inferred $B_\parallel$ at \TAUE{-6.0}, $-5.0$, and $-4.0$ (three left panels, 
  from left to right, respectively).
  The rightmost panel shows the height where the plasma $\beta=1$ (assuming 
  the magnetic field is parallel to the \gls*{los}). 
  Contour curves are the same as in \fref{fig6}.}
  \label{fig8}
\end{figure*}

\begin{figure*}[htp]
  \center
  \includegraphics[width=0.9\textwidth]{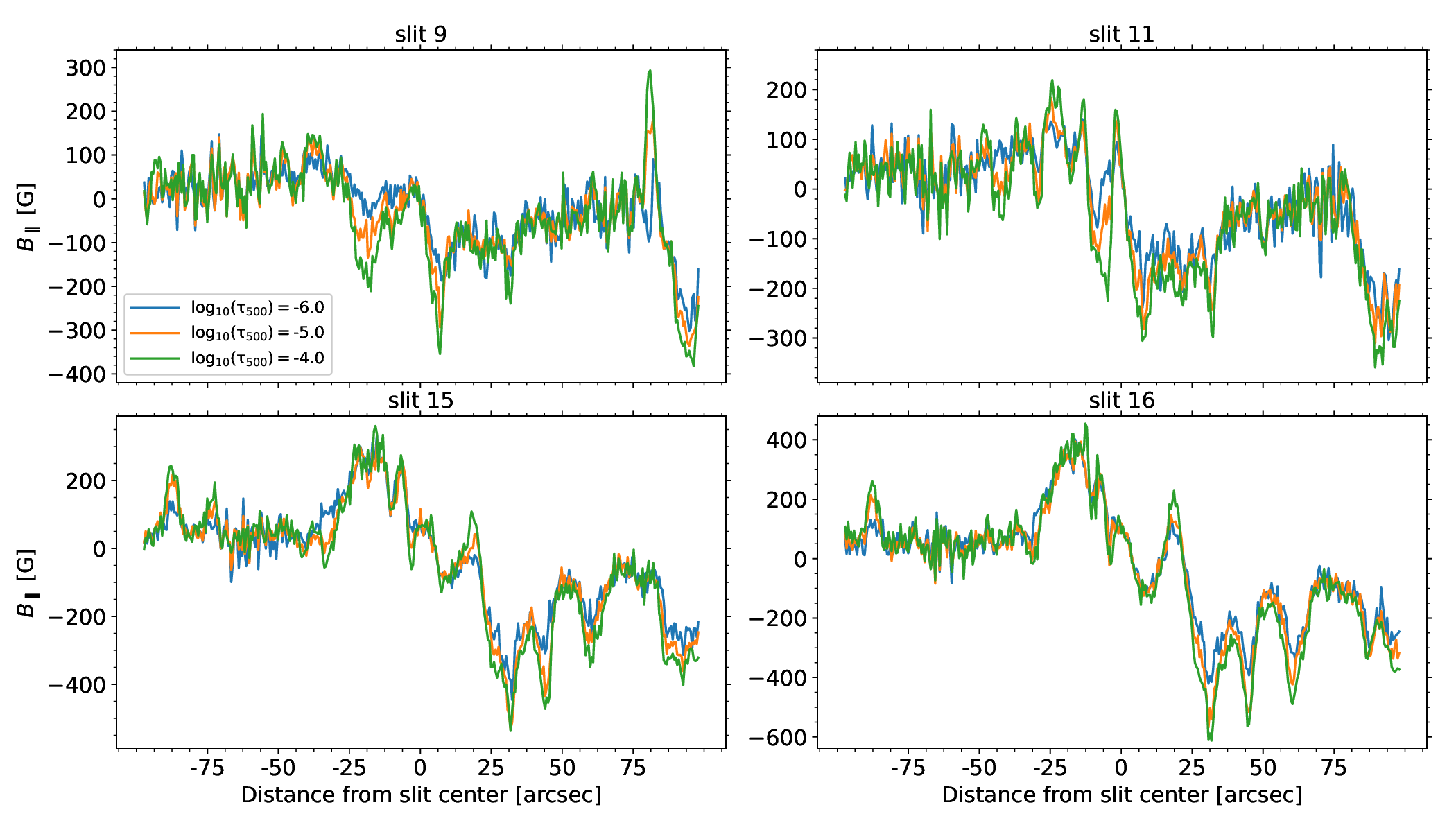}
  \caption{Inferred {\blos} at \TAUE{-6.0} (blue curves), $-5.0$ (orange curves), and
  $-4.0$ (green curves) for slit locations 9, 11, 15, and 16 (see label on top of
  the panels). The slit positions on the observation field of view are indicated
  by the blue dashed curves in \fref{fig1}.}
  \label{fig9}
\end{figure*}

We applied an inversion strategy similar to that outlined in \citet{Li2023ApJ}, namely
two non-magnetic cycles to obtain the thermodynamic model, i.e., 
temperature ($T$), \gls*{los} velocity ($v_\parallel$), micro-turbulent velocity 
($v_{\rm turb}$), and gas pressure ($P_{\rm g}$), from just the intensity profile. 
The model atmosphere used in the spectral synthesis is stratified with 60 non-equally 
spaced layers between \TAUE{-8.0} and 1.0. In order to reduce the significant 
computational requirements, once the Stokes $I$ inversion is completed, we fix 
the thermodynamic quantities of the model and we only invert the longitudinal 
magnetic field ($B_\parallel$) from the observed circular polarization in two cycles. 
As in \citet{Li2023ApJ}, the plasma velocity and the magnetic field 
are assumed to be parallel to the local vertical in the inversion, since the 
non-axial symmetric components of the velocity and the magnetic field 
significantly increase the computing time, without significantly impacting the
circular polarization for the \gls*{los} of the observation. The vertical
components ($v_z$ and $B_z$) are then projected onto
the \gls*{los}, and these longitudinal components are the ones constrained by
the observation. In Table \ref{tab1} we show
the number of nodes for each variable in each cycle. 
In addition, it is only in the last cycle that we take the radiation field anisotropy 
into account. In this paper we only show the nodes of the inversion between \TAUE{-6.8} and
$-2.5$. During the inversion we also considered two nodes at \TAUE{-8.0} and $1.0$,
at the top and bottom boundaries of the model atmosphere, respectively. Moreover,
for the temperature we considered an additional node at around \TAUE{-1.8}.
This node selection was the result of 
experimentation with the inversion of the intensity profiles.

The errors in the inferred parameters are computed from the diagonal of the Hessian
matrix \citep[e.g.][]{delToroIniesta2003}. The uncertainties given by this method
indicate how well contrained a node value is relative to the others \citep{Milic2018A&A}.
One of the best ways to estimate the confidence interval of the inferred node values is 
via Bayesian inference by implementing a Markov chain Monte Carlo
\citep[e.g.,][]{AsensioRamos2007A&A,Li2019ApJ}. However,
more than $10^5$ synthesis are typically required to achieve a good posterior 
distribution, and thus this approach is suitable for very fast forward models. 
Besides these two methods, Monte Carlo simulations have also been used to 
estimate uncertainties \citep{WestendorpPlaza2001ApJ,SainzDalda2023ApJ}. 
In this method random noise is added to the Stokes profiles, and the standard
deviation of the results of the inversion of these profiles is representative
of the uncertainty. An example of the uncertainty estimated using this method
is shown in Sec.~\ref{sec4.3}. Note that all the uncertainties mentioned above
are representative of how the inferred parameters can change without significantly
impacting the merit function of the inversion, i.e., the measurement of the
goodness of the fit.

In the spectral region observed by the \gls*{clasp2.1} there are three resonance
lines of \ion{Mn}{1} at 279.56, 279.91, and 280.19~nm, as well as two blended lines
of \ion{Mg}{2} at 279.88~nm. Including these transitions in the inversion helps
contraining the inference in the upper photosphere and lower chromosphere. However,
we have found that, for these data, the results of the inversion of the \ion{Mg}{2}
h and k lines including and neglecting these other lines are compatible within the
inversion errors. Therefore, to ease the already significant computational cost, we
have performed the inversions pixel by pixel using a \ion{Mg}{2} model atom with four
levels, the lower and upper levels of the \ion{Mg}{2} h and k lines and the \ion{Mg}{3}
ground level. We show the comparison between the inversions including and neglecting
the \ion{Mg}{2} UV triplet and the \ion{Mn}{1} resonance lines in Appendix~\ref{appendix}.

\fref{fig5} shows observations of the Stokes 
$I$ and $V$ signals at the wavelengths indicated in the caption and the fits 
resulting from the inversions. Overall the fits present the main features of 
the observation both at the line center and in the wings. The fits of the 
circular polarization are smoother than the observation, which is
expected due to the polarization noise of the observation.


\section{Results}\label{sec4}

In this section we show the results of the inversion of the \gls*{clasp2.1}
data following the strategy described in Sec.~\ref{sec3}. In particular
we emphasize the inversion results in the plage region (Sec.~\ref{sec4.1}),
the penumbra and superpenumbra (Sec.~\ref{sec4.2}), and in a region where we
find a change of the magnetic field polarity with height (Sec.~\ref{sec4.3}).

\subsection{The plage and the overlying moss}\label{sec4.1}

The plage region observed by the \gls*{clasp2.1} suborbital experiment, approximately 
covering the region between 10 -- 60~arcsec in the X direction and 20 -- 100~arcsec 
in the Y direction (see \fref{fig1}) shows strong emission features in the 
AIA 1600 \AA\ and IRIS 2796 \AA\ bands (panels (C) and (D), respectively). 
The dominant magnetic flux is negative in the underlying photosphere (panel (E)).
From the HMI magnetogram, the longitudinal component of the magnetic
field in these photospheric flux concentrations is about $-400$~G on average, 
reaching about $-1000$~G in some flux concentrations\footnote{These field strengths 
are somewhat smaller than those derived from the 
observations with the SOT/SP instrument onboard the  
Hinode satellite, because of the assumption of filling factor unity in the HMI 
inversions.}.
A moss region over the weakest part of the plage, between 
10 -- 60 ~arcsec in the X direction and
20 -- 100~arcsec in the Y direction, close to
the footpoints of the hot coronal loops, shows bright emission features in the 
304 and 171 \AA\ AIA bands \citep[panels (A) and (B),][]{Berger1999SoPh}. 
The moss is a hot layer in the transition region with a temperature of about 
1~MK \citep{Martens2000ApJ}.

An intensity and circular polarization profile representative of those found in
the plage region is shown in \fref{fig4}, corresponding to the
red ``$\times$'' symbol in panel (E) of \fref{fig1}. The intensity profiles of the 
\ion{Mg}{2} h and k lines show almost no central reversal. Consequently, the circular
polarization profiles only show two lobes (because the circular polarization profile
resembles the first derivative of the intensity). 
The inferred temperature stratification in the plage region 
has a temperature of about 6000~K in the middle chromosphere, in agreement
with the results of \citet{Carlsson2015ApJ}.
The inferred longitudinal magnetic field is about $-500$~G at \TAUE{-4.5}, 
corresponding to the middle chromosphere, and
about $-400$~G at \TAUE{-6.0}, corresponding to the upper chromosphere.

The purple contours in panels (A) and (K) of \fref{fig6} correspond to 
the brightness in the 171 \AA\ band intensity image. 
The shape of the hot moss region roughly matches the region
at \TAUE{-6.0} with larger $T$ and $n_{\rm e}$ and relatively larger
$v_{\rm turb}$. 
Larger gas pressure is required to fit the profiles in the plage region, 
in agreement with the results in \citet{delaCruz2016ApJ}.
The inferred $n_{\rm e}$ and $P_{\rm g}$ in this region are
$\rm 1.9\times10^{11}$ -- $\rm 2.4\times10^{12}~cm^{-3}$ and 
0.5 -- 6.7 $\rm dyn\cdot cm^{-2}$, respectively. 
The gas pressure that we infer, 2.5 $\rm dyn\cdot cm^{-2}$ on average,
is slightly larger than the values in a
moss region estimated with a differential emission measure analysis by
\citet{Fletcher1999ApJ}, namely 0.7 -- 1.7 $\rm dyn\cdot cm^{-2}$.
This is reasonable since the moss is located in the transition region above the 
formation region of the line centers of the h and k lines.
Note that in our inversion we assume both hydrostatic equilibrium and ionization
fractions in local thermodynamic equilibrium in the equation of state.
Despite these assumptions being relatively common in many inversion applications, 
they introduce an additional and difficult to quantify uncertainty in the 
inferred $P_{\rm g}$ and $n_{\rm e}$. Finally, in this region $v_\parallel$ is 
downflowing with respect to the lower chromosphere.

\begin{figure*}[htp]
  \center
  \includegraphics[width=0.9\textwidth]{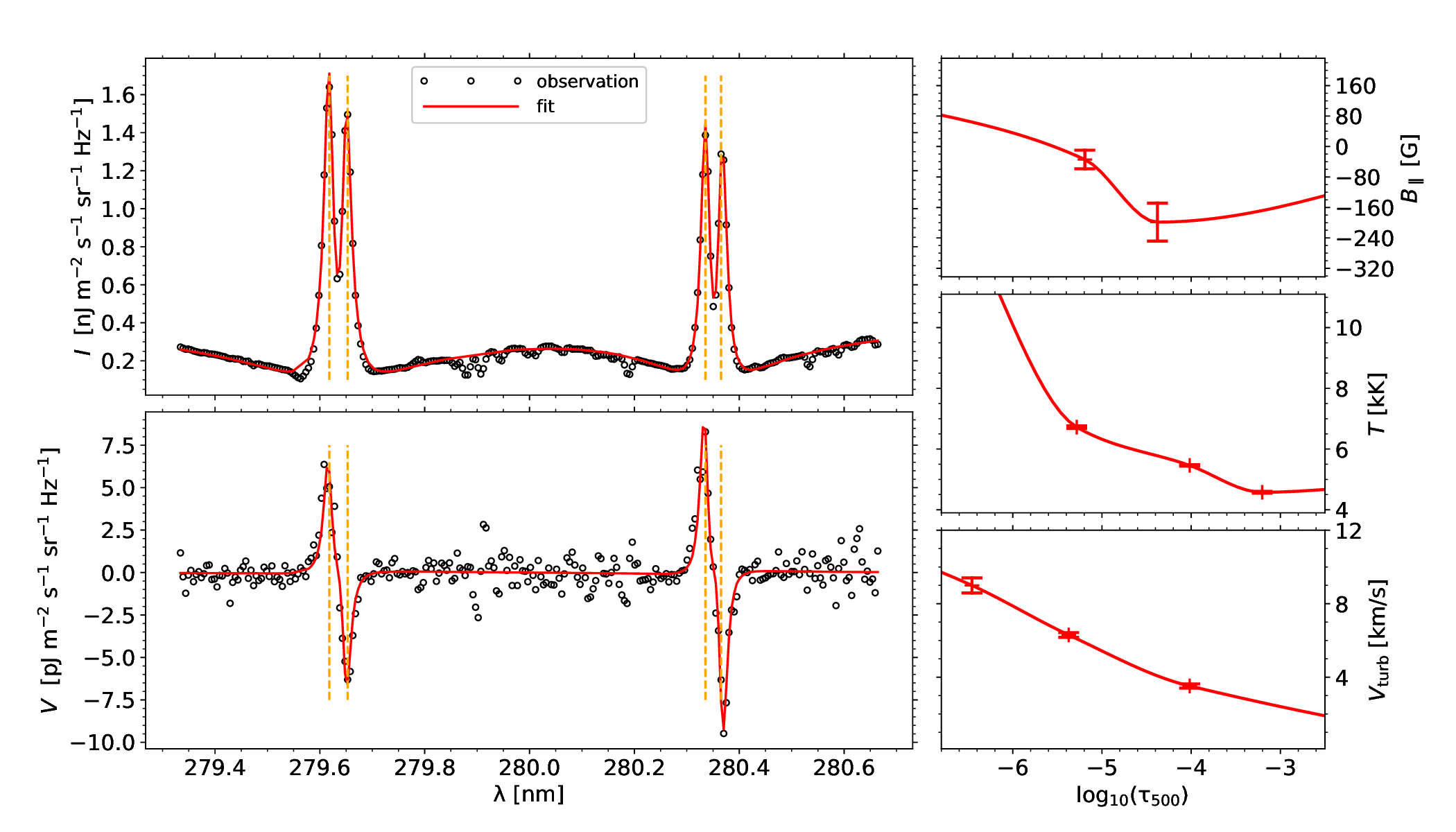}
  \caption{ Same as \fref{fig4}, but for a pixel in the vicinity of the penumbra
  (yellow ``$\times$'' symbol in Panel (E) of \fref{fig1}).
  The vertical orange dashed lines indicate the location of the
  ${\rm k_2}$ and ${\rm h_2}$ peaks.}
  \label{fig2}
\end{figure*}

In the underlying layer at \TAUE{-4.5}, the inversion also shows an increase of
both $T$ and $n_{\rm e}$ in the plage region underneath the moss, 
with respect to the quiet region. These hot regions are distributed similarly 
to the regions with larger magnetic flux in the photosphere
(see the black and blue contours in panel (B) of \fref{fig6}),
but covering an expanded area with respect to the latter. Inside these regions
with larger photospheric magnetic flux,
the inferred $n_{\rm e}$ is also larger (see the white arrows in panel (F) 
of \fref{fig6}), 
which suggests that the chromospheric heating 
is more significant inside the magnetic flux concentrations. 
Panel (A) of \fref{fig7} shows a scatterplot between $n_{\rm e}$ at 
\TAUE{-4.5} and the HMI longitudinal magnetic fields for the region delimited
by the solid purple lines in panel (B) of \fref{fig6}. The correlation 
coefficient between both
quantities is about $0.6$, which indicates the existence of some correlation, even
if not a strong one.
In the scatterplot, the majority of the points are located in the lower 
left corner (longitudinal magnetic field of 100~G or less). The
reason is that, in the photosphere, most of the field of view does not show
strong longitudinal magnetic field amplitudes and the inferred electron density is
small in most of such areas. Besides, the region with larger electron density covers
a larger area with respect to the photospheric longitudinal magnetic field, resulting in
a number of points in the scatterplot with relatively large electron density but
weak photospheric longitudinal magnetic field.

In \fref{fig8} we show the inferred $B_\parallel$ at \TAUE{-6.0}, $-5.0$, and $-4.0$ 
(first, second, and third panels, from left to right, respectively). The contours 
in \fref{fig8} are the same as in \fref{fig6}. 
The inferred $B_\parallel$ in the regions with photospheric flux concentrations 
decreases with height. The polarity of the inferred magnetic fields is
consistent with that of the underlying photospheric magnetic fields. The strongest 
magnetic fields are found at the same location as the photospheric magnetic flux 
concentrations. The area occupied by the magnetic field concentrations is larger 
in the chromosphere than in the photosphere; these concentrations have already 
expanded at a height of \TAUE{-4.0}, almost filling up the apparently unmagnetized 
gaps between the flux concentrations observed in the photosphere. This is in 
agreement with the conclusion by \citet{Morosin2020A&A}, which posit that the magnetic
canopy has formed in the lower chromosphere, where the \ion{Mg}{1} line
at 5172 {\AA} forms. The magnetic field strengths {in the chromosphere} reach about 
$-100$~G between the photospheric flux concentrations, and about $-600$~G inside them.

The rightmost panel of \fref{fig8} shows the optical depth, $\log_{10}\tau_{500}$,
where $\beta=1$ (with $\beta=2\mu_0P_g/B_{\parallel}^2$, under the assumption 
that the magnetic field is parallel to the \gls*{los}, with $\mu_0$ the magnetic 
permeability). Above the regions where we find
the photospheric magnetic flux concentrations, the magnetic pressure
overcomes the gas pressure in the lower chromosphere.
This region of the atmosphere is below the
main sensitivity range of the circular polarization profile of the \ion{Mg}{2} 
h and k lines.

Panels (B) and (C) of \fref{fig7} show scatterplots between $B_\parallel$ and 
$n_{\rm e}$ and  between $B_\parallel$ and $T$  at \TAUE{-4.5}, respectively. 
The correlation coefficient is about $0.8$, a correlation indicative of the magnetic 
origin of the heating in the plage, in agreement with the result of 
\citet{Ishikawa2021SA}.

The correlation between the intensities in the moss region and the 
underlying chromosphere, e.g. between the 171~{\AA} and $\rm Ly_\alpha$ intensities
\citep{Vourlidas2001ApJ}, between the 171~{\AA} and $\rm H_\alpha$ intensities
\citep{DePontieu2003ApJ}, or between the 172~{\AA} and 2796~{\AA} intensities
\citep{Bose2024NatAs} suggests that the heating mechanism for both the
chromosphere plage and the overlying moss are closely related. 
The structures in the inferred magnetic field map at \TAUE{-6.0} and $-4.5$ 
shown in \fref{fig8} are morphologically similar to the outline of the moss region. 
Panels (D) -- (G) of \fref{fig7} show scatterplots between the intensity
in the AIA 171~{\AA} band and the indicated quantities resulting from the inversion.
The correlation coefficients  of
about $0.8$ between the AIA 171~{\AA} intensity and
the (inferred) longitudinal magnetic fields at \TAUE{-6.0} and $-4.5$ 
suggest not only a magnetic origin for the heating in the
chromosphere of the plage region, but also for the heating of the moss region in the
transition region.

\begin{figure*}[htp]
  \center
  \includegraphics[width=0.9\textwidth]{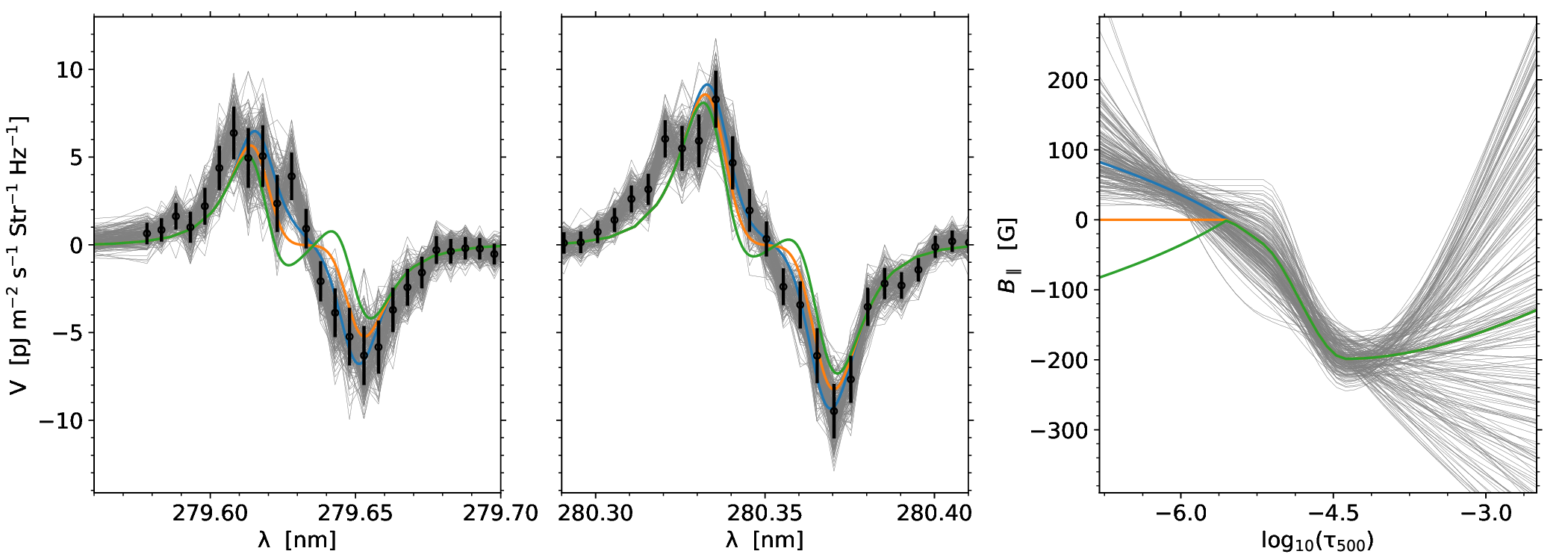}
  \caption{Two left panels: observed (open circles with error bars) and synthesized 
  (blue, orange, and green curves) circular polarization proﬁles for the pixel 
  in the vicinity of the penumbra considered in \fref{fig2}. Gray curves show 200 
  Stokes $V$ profiles generated by adding 
  random noise to the observation.
  Right panel: blue (the magnetic field model from the inversion), orange (with zero 
  magnetic fields in the upper chromosphere) and  green (with flipped magnetic 
  fields in the upper chromosphere) curves show the corresponding magnetic fields.
  Gray curves show the 200 Stokes $V$ profiles mentioned in the text.}
  \label{fig10}
\end{figure*}


\subsection{The penumbra and superpenumbra} \label{sec4.2}

In panel (D) of \fref{fig1}, within the region spanning 30 -- 45~arcsec in the 
X direction and $-50$ -- 0~arcsec in Y direction, the IRIS 2796 \AA\ 
slit-jaw image shows elongated fibril structures
outward the penumbra, corresponding to the superpenumbra \citep{Loughhead1968SoPh}, 
which is more distinctly observed in $\rm H_{\alpha}$ and in the \ion{He}{1} triplet 
lines at 10830~{\AA} \citep{Schad2013ApJ}. 
This specific region is covered by slits 8 -- 13 of the \gls*{clasp2.1} observation, 
which presents such fibrils in the intensity image at $\rm k_{2b}$ (top left panel 
of \fref{fig5}). The underlying photosphere in this region exhibits mixed polarities
(see the region delimited by the dashed purple lines in panel (B) of 
\fref{fig6}). Bright features can be seen in the AIA 1600 \AA\ band
and IRIS 2796 \AA\ slit-jaw images in these regions, while in the AIA 304 \AA\
band there is a lack of bright features, with the exception of some bright
fibrils (see \fref{fig1}). The inversion results show a larger temperature 
at \TAUE{-6.0} and $-4.5$ in the superpenumbra region compared to the 
quiet regions, in agreement with the results of \cite{SainzDalda2019ApJ}. 
In the penumbra, which occupies the region between about $-25$ -- 0~arcsec for slits
15 and 16, the temperature is not significantly enhanced with respect to other regions.

At \TAUE{-6.0}, the inversion returns a $P_{\rm g}$ of less than 
1~$\rm dyn\cdot cm^{-2}$, much lower than the average value inferred in 
the plage region. There is no remarkable increase in $n_{\rm e}$ either.
However, some fibrils can be seen in the $n_{\rm e}$, $v_\parallel$,
$v_{\rm turb}$, and $P_{\rm g}$ maps (see upper panels of \fref{fig6}). 

The inferred $B_\parallel$ at {\TAUE{-4.0}} shown in \fref{fig8} also
exhibits mixed polarities in the region between 30 -- 45~arcsec in the X direction
and  between slits 8 -- 13, and the polarity is
the same as in the photosphere. The inferred $B_\parallel$ decreases its amplitude
with height. At {\TAUE{-6.0}}, part of the negative flux
disappears, for instance at around $-45$~arcsec in slits 13 and 14. As shown in
\fref{fig6}, the regions with larger $T$ and $n_{\rm e}$ are outside the penumbra.
The overall distribution of the regions with larger $T$ is similar to the distribution
of the regions with larger magnetic flux. This suggests that the heating in the
chromosphere of the vicinity of the sunspot is also of magnetic 
origin. However, when we focus on
smaller scales, contrary to what was found for the
plage region, the larger $n_{\rm e}$ and $T$ areas are usually
located between magnetic flux concentrations (see the green arrows in 
panel (F) of \fref{fig6}). Panels (G) and (I) of \fref{fig7} present scatterplots between 
$B_\parallel$ and $T$, and between $B_\parallel$ and $n_{\rm e}$ at \TAUE{-4.5} 
for the region delimited by the dashed purple lines in \fref{fig6}. The corresponding
correlation coefficients are about $-0.3$, a relatively weak negative correlation
(note, however, the relatively small size of the available sample),
significantly different to the correlations found for the plage region.
Therefore, the details of the heating mechanisms in this region, even if
the magnetic field still plays a significant role, may differ with
respect to those in the plage region. It is important to emphasize that in this paper 
we have focused on inferring the longitudinal component of the magnetic field,
so it is likely that these hot regions in between magnetic flux regions are also
magnetized, but with a magnetic field which is significantly inclined with
respect to the \gls*{los}.

\subsection{Change of the magnetic field polarity with height}\label{sec4.3}

The blue, orange, and green curves in \fref{fig9} show $B_\parallel$ at
\TAUE{-6.0}, $-5.0$, and $-4.0$, respectively, for slits 9, 11, 15, and 16
of the \gls*{clasp2.1} observation (with the corresponding slit indicated
on top of each panel). The slits locations are indicated by the blue dashed lines
in \fref{fig1}. Slit 16 is the rightmost slit, which crosses the
edge of the penumbra and the central region of some flux concentrations in
the plage region.

In slit 16 (bottom right panel of \fref{fig9}) $B_\parallel$
reaches about $-600$~G at \TAUE{-4.0} in the plage region, decreasing to about
$-400$~G at \TAUE{-6.0}. At the edge of the penumbra there is an
apparent lack of variation in $B_\parallel$. 

Consistent with the weak field approximation analysis by \citet{Ishikawaforthcoming}, 
the inversion results also indicate changes in the polarity of the longitudinal 
magnetic field component $B_\parallel$ with height in certain areas of the slits 9, 
11, and 15. For instance, such polarity changes are detected in the region between 
77 and 82~arcsec in slit 9, and in the region between $-5$ and 0~arcsec in slit 11 
(see \fref{fig9}). In \fref{fig2} we show Stokes profiles corresponding
to a pixel with such a polarity change. The intensity profile shows a significant
reversal in the core of the \ion{Mg}{2} h and k lines. The orange dotted lines in 
the figure indicate the ${\rm k_2}$ and ${\rm h_2}$ peaks. For an intensity profile with this 
shape, the circular polarization profile is expected to show either two pairs of 
lobes (see, e.g., the profile in Appendix~\ref{appendix} corresponding to the green
``$\times$'' symbol in \fref{fig1}) or almost zero inner lobes if $B_\parallel$ is
too weak \citep{Li2023ApJ}. However, \fref{fig2} shows both inner and outer lobes in
circular polarization, with the same sign, thus resembling a two-lobe profile.
For this to happen a polarity change from the middle to the upper chromosphere is
necessary. The inversion of this profile returns a negative $B_\parallel$ polarity 
at \TAUE{-4.5} (about $-200$~G) and a small, albeit positive $B_\parallel$ polarity, 
at \TAUE{-6.0} (about 40~G). 
Since the error estimatted from the diagonal of the Hessian matrix is larger 
than 40~G, we carried out a Monte Carlo simulation to validate the polarity change 
\citep{WestendorpPlaza2001ApJ,SainzDalda2023ApJ}. 
We generated 200 profiles by adding random Gaussian noise to the observation.
The gray curves in the two left panels of \fref{fig10} show the 200 Stokes $V$ profiles. 
We then applied the HanleRT-TIC to all profiles. Since we only use this method to
investigate the uncertainty in $B_\parallel$, the thermodynamic parameters are fixed
and only $B_\parallel$ is retrieved in one cycle with four nodes.
The node values are randomly initialized between -500 and 500~G.
The inferred $B_\parallel$ are shown in the right panel of \fref{fig10} (see gray
curves). As it was expected, all 200 inversions confirm the polarity change with height.
We want to emphasize that this change in polarity was expected because it is a
necessary condition for the observed shape of the circular polarization profile. 
To further demonstrate the need for this change of polarity, we have synthesized
the circular polarization profiles for stratifications of $B_\parallel$ with both
zero field and with no change of polarity in the upper chromosphere (see \fref{fig10}).
As expected, if there is no polarity change the circular polarization profile
shows a four-lobe shape, while the zero magnetc field shows signals very close to
zero at the center of the circular polarization profile. Both syntheses fail to
fit the shape of Stokes $V$ close to the line center, further confirming the
polarity change.


\section{Summary and conclusions}\label{sec5}

The HanleRT-TIC has been applied to the spectro-polarimetric 
observation of the \ion{Mg}{2} h and k lines obtained by the \gls*{clasp2.1} 
suborbital space experiment. 
The observation, consisting of a slit scan, covers a plage region and the
edge of a sunspot penumbra.
We have obtained a map of the magnetic field longitudinal component 
in the upper chromosphere and the stratification of the thermodynamic model,
including the temperature, \gls*{los} velocity,
micro-turbulent velocity, and electron density and gas pressure.

In agreement with previous studies based on intensity observations of the
\ion{Mg}{2} h and k lines with IRIS, the inverted models show larger 
$T$ and $n_{\rm e}$ in the plage region \citep{delaCruz2016ApJ,SainzDalda2019ApJ}.
The observed plage is dominated by negative (pointing toward the solar surface) 
magnetic flux in both the photosphere and the middle and upper chromosphere. 
The inferred $B_\parallel$ 
in the chromosphere covers an area clearly expanded with respect to the magnetic 
field concentrations of the
photospheric magnetogram, almost filling the unmagnetized gaps between the 
flux concentrations, which indicates that the magnetic field expands and
fills the chromospheric volume below the middle chromosphere, either 
in the lower chromopshere \citep{Morosin2020A&A} 
or in the photosphere \citep{Buehler2015A&A}. 
The pattern of the inferred
$B_\parallel$ map overall matches the larger values of $T$ and $n_{\rm e}$
in the maps in 
the chromosphere, as well as the moss region in the overlying transition 
region seen in the AIA 171~{\AA} band.
Apart from the correlation between the intensity of the 171~{\AA} band 
and $T$ and $n_{\rm e}$ \citep{Bose2024NatAs}, 
in this work we also show the correlation between the 171~{\AA} intensity
and $B_\parallel$ in the middle and upper chromosphere, which suggests the 
magnetic origin of the heating in both the plage and moss regions.
Such correlation does not exist between the 171~{\AA} intensity and the magnetic 
field inferred from the \ion{Ca}{2} line at 854.2~nm with the \gls*{wfa} 
\citep{Judge2024ApJ}, possiblly due to the lower formation height of \ion{Ca}{2} 
854.2~nm with respect to the \ion{Mg}{2} h and k lines.
Moreover, the correlation between $B_\parallel$ and $T$ and $n_{\rm e}$ 
in the chromomosphere also suggests that heating is more significant inside the 
magnetic field concentrations. 
We also find a moderate correlation between $n_{\rm e}$ in 
the middle chromosphere and $B_\parallel$ in the photosphere, which can also 
be seen by visual inspection of panel (F) of \fref{fig6}.
Such a correlation is different to that found in previous studies such as that
by \citet{Anan2021ApJ}, reporting
no significant correlation between the magnetic field inferred
from the \ion{He}{1} triplet at 1083.0~nm and the energy flux is found.

In agreement with \citet{SainzDalda2019ApJ}, 
the inferred $T$ and $n_{\rm e}$ in the superpenumbra region also show
an increase with respect to the quiet regions. 
This increase is clearly related to the inferred magnetic field.
However, their larger values are usually found between
the magnetic field concentrations. 
The weak negative correlation between $B_\parallel$ and both $T$ and 
$n_{\rm e}$ is completely different from the correlations for the plage region.
This suggests that, while the heating in this region should be of magnetic origin,
the details or particularities of the heating mechanism may show differences with
respect to those in the plage and moss regions. It is of interest to emphasize that 
in this paper we have focused on inferring only the longitudinal component of the 
magnetic field. 
Significantly inclined magnetic fields with respect to the \gls*{los} along
the superpenumbral fibrils have been reported by \citet{Schad2013ApJ} from
spectropolarimetric observation of the \ion{He}{1} 1083.0~nm multiplet.
Therefore, these hot regions are likely filled with more inclined magnetic
fields.

Generally, we find that the inferred $B_\parallel$ decreases with height 
in the middle and upper chromomosphere. 
In the plage region $B_\parallel$ can still reach about $-400$~G in the 
upper chromosphere, which is consistent with the field strengths  
inferred from observations of the \ion{Ca}{2} 854.2~nm line
\citep{Pietrow2020A&A,Morosin2020A&A,daSilvaSantos2023ApJ} 
and the \ion{He}{1} triplet at 1083.0~nm \citep{Anan2021ApJ}.
In the penumbra, $B_\parallel$
does not show a significant variation with height, although there could 
still be some variation
compatible with the uncertainties and the polarization noise. 
It is also noteworthy that, in agreement with the weak field approximation analysis by
\citet{Ishikawaforthcoming}, our inversion results reveal  
changes with height in the magnetic field polarity in certain regions, 
namely near the penumbra and in the pore. Such a polarity change with height can be
explained with a magnetic field configuration in which the magnetic field, which is
anchored to the sunspot, bends down toward the photosphere 
at different locations for different heights, resulting 
in a change of polarity for a number of lines of sight.

Both, CLASP2 and CLASP2.1 measured the wavelength variation of the four Stokes
parameters, but in this and in our previous paper we have focused on the inversion of the 
Stokes $I$ and $V$ profiles, which have allowed us to infer  
the longitudinal component of the magnetic field. In forthcoming papers  
we will consider the full Stokes-vector inversion problem of the \ion{Mg}{2} h and k lines,
showing inference results for the quiet and plage regions observed by these  
novel suborbital space experiments.   

\acknowledgements
We gratefully acknowledge the financial support from the European Research Council (ERC) 
under the European Union's Horizon 2020 research and innovation programme (Advanced 
Grant agreement No. 742265). T.P.A.'s participation in the publication is part 
of the Project RYC2021-034006-I, funded by MICIN/AEI/10.13039/501100011033, and 
the European Union “NextGenerationEU”/RTRP. 
TdPA and JTB acknowledge support from the Agencia Estatal
de Investigación del Ministerio de Ciencia, Innovación y Universidades (MCIU/AEI)
under grant ``Polarimetric Inference of Magnetic Fields'' and the European
Regional Development Fund (ERDF) with reference
PID2022-136563NB-I00/10.13039/501100011033.
L.B. and J.T.B. gratefully 
acknowledge the Swiss National 
Science Foundation (SNSF) for financial support through grant CRSII5\_180238. 
CLASP2.1 is 
an international partnership between 
NASA/MSFC, NAOJ, JAXA, IAC, and IAS; additional partners 
include ASCR, IRSOL, LMSAL, and the University of Oslo. 
The Japanese participation was funded 
by JAXA as a Small Mission-of-Opportunity Program, 
JSPS KAKENHI Grant numbers JP25220703 
and JP16H03963, 2015 ISAS Grant for 
Promoting International Mission Collaboration, and by 2016 
NAOJ Grant for Development Collaboration. 
The USA participation was funded by NASA Award 
16-HTIDS16\_2-0027. 
The Spainish participation was funded by the European Research Council through 
Advanced Grant agreement No. 742265. 
The French hardware participation was funded by 
CNES funds CLASP2-13616A and 13617A. AIA and HMI data are courtesy of NASA/SDO 
and the AIA, and HMI science teams. IRIS is a NASA small explorer mission developed and 
operated by LMSAL with mission operations executed 
at NASA Ames Research Center and major 
contributions to downlink communications funded by ESA and the Norwegian Space Centre.

\appendix
\twocolumngrid

\begin{figure*}[htp]
  \center
  \includegraphics[width=0.9\textwidth]{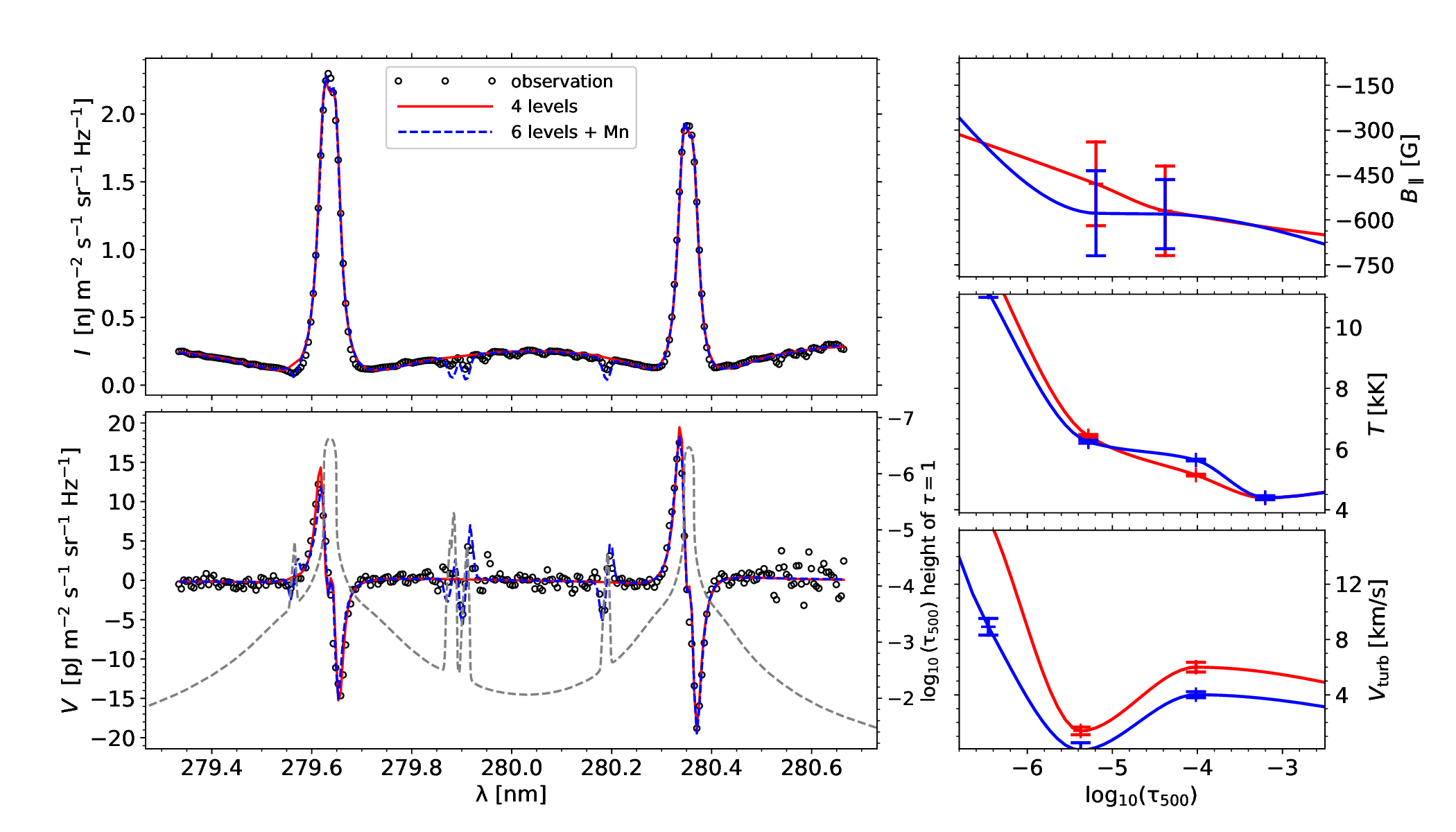}
  \caption{ Same as \fref{fig4} but including the inversion accounting for
  the \ion{Mg}{2} UV triplet and the \ion{Mn}{1} lines at 279.56, 279.91,
  and 280.19~nm. The dashed gray curve in the bottom left panel shows the
  optical depth (in $\rm log_{10}(\tau_{500})$) corresponding to the 
  atmosphere height for which the optical depth at any given wavelength
  is equal to one ($\tau_\lambda=1$).}
  \label{fig4a}
\end{figure*}

\begin{figure*}[htp]
  \center
  \includegraphics[width=0.9\textwidth]{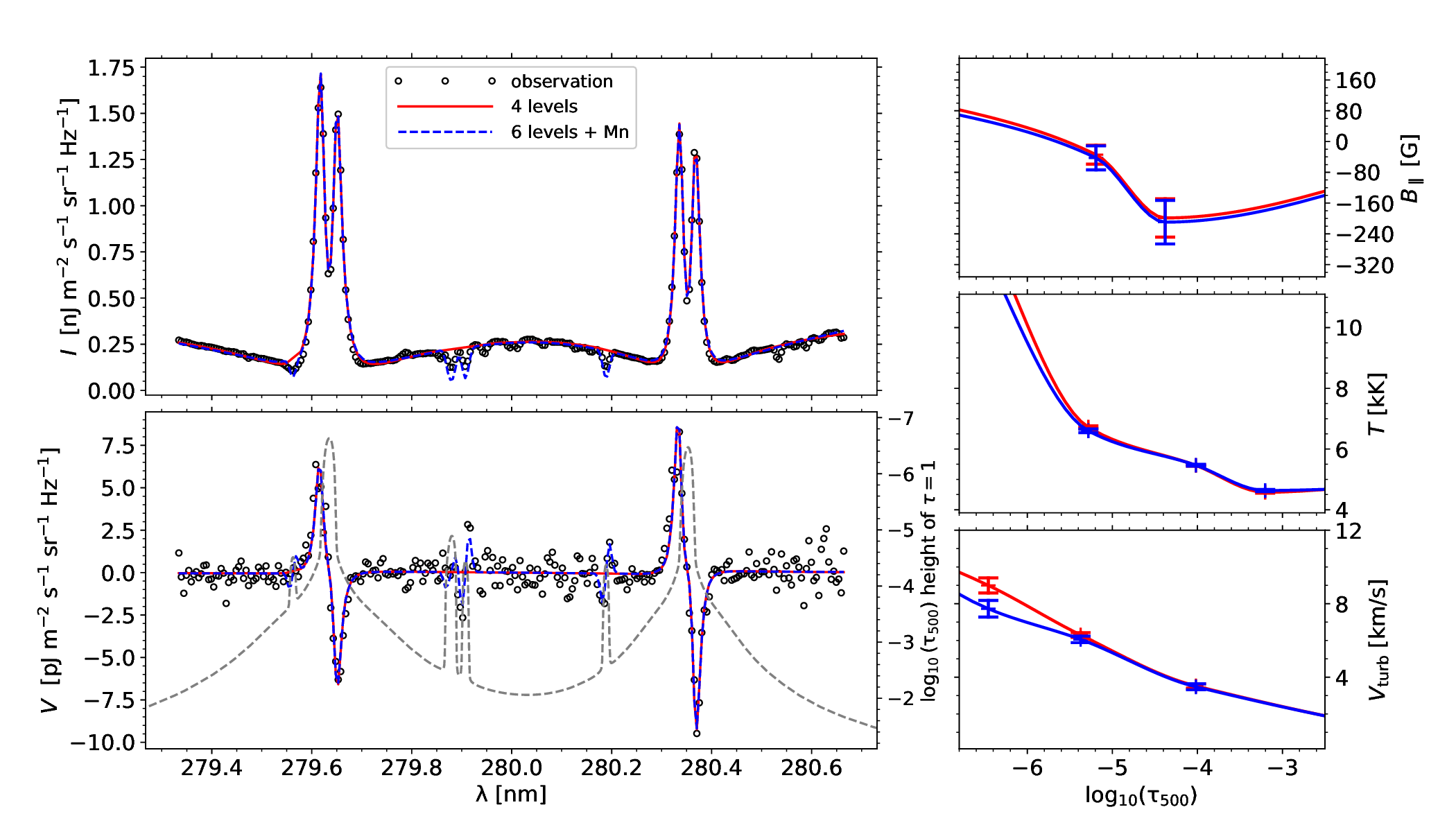}
  \caption{Same as \fref{fig4a} but for a pixel in the vicinity of the penumbra
  (yellow ``$\times$'' symbol in Panel (E) of \fref{fig1}). The red solid curve is
  the same as in \fref{fig2}.}
  \label{fig2a}
\end{figure*}

\begin{figure*}[htp]
  \center
  \includegraphics[width=0.9\textwidth]{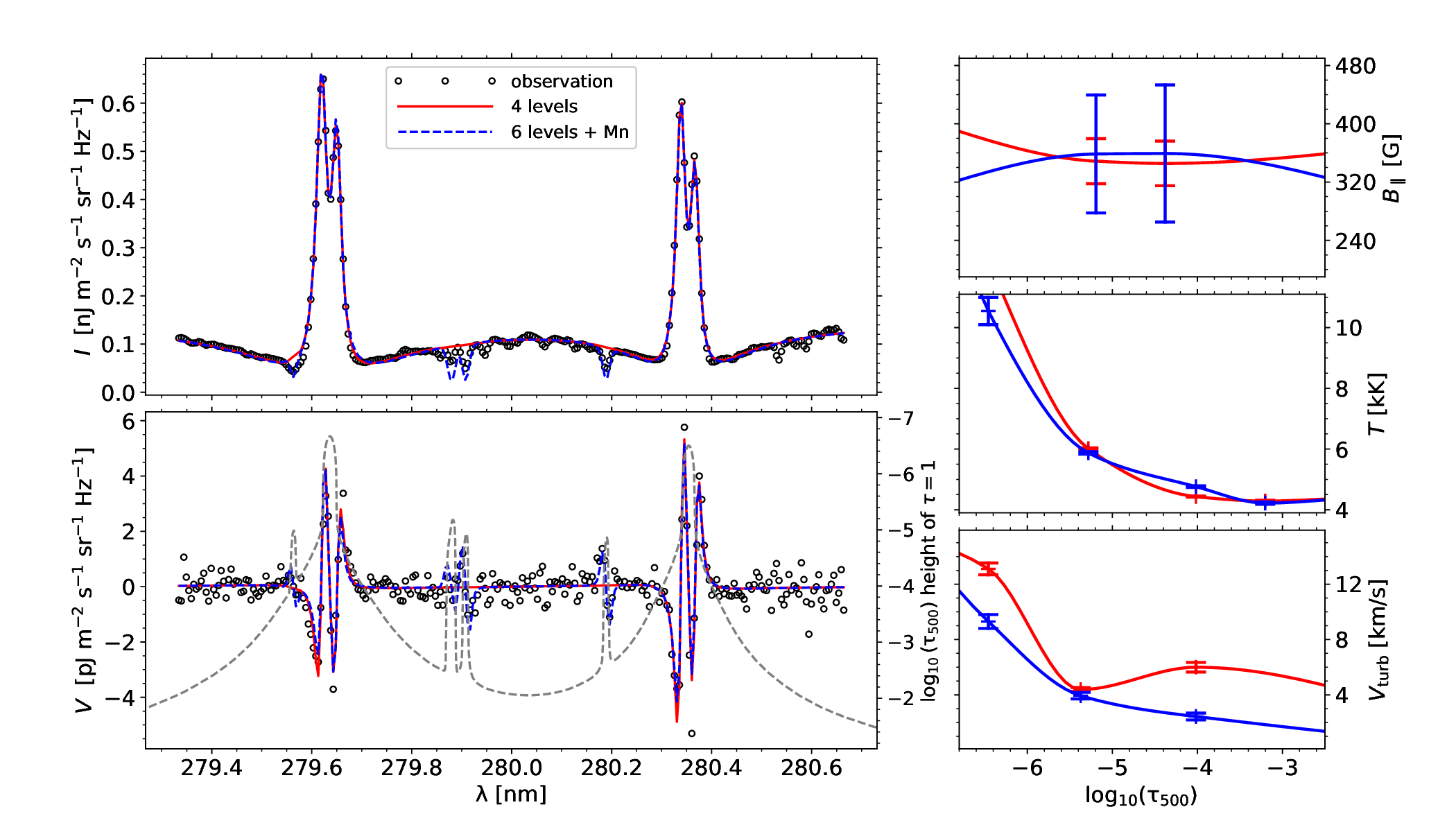}
  \caption{Same as \fref{fig4a}, but for a pixel in the penumbra 
  (green ``$\times$'' symbol in Panel (E) of \fref{fig1}).}
  \label{fig3a}
\end{figure*}

\section{The impact of the \ion{Mn}{1} lines and the \ion{Mg}{2} subordinate 
lines on the inferred model}\label{appendix}

In the spectral region observed by the \gls*{clasp2.1} there are three resonance
lines of \ion{Mn}{1} at 279.56, 279.91, and 280.19~nm, 
as well as two blended lines of \ion{Mg}{2} at 279.88~nm. These lines form in
the lower chromosphere \citep{Pereira2015ApJ,Tanausu2020ApJ,Tanausu2022ApJ}. 
Here we investigate the impact of including these lines in the inversion. Even
though accounting for HFS is necessary to correctly model the \ion{Mn}{1} lines, its general 
treatment is not included in our inversion code. Consequently, we neglect HFS in the
inversion. Neglecting the HFS affects the width of the lines and leads to an
underestimation of the circular polarization signal \citep{Tanausu2022ApJ}.
Moreover, their accurate modeling requires an atomic model with a large number of levels
and transitions. Due to these reasons, our modeling of the \ion{Mn}{1} profiles is
very approximate. Nevertheless, including the \ion{Mn}{1} lines and the \ion{Mg}{2}
UV triplet lines while giving more weight in the inversion to the \ion{Mg}{2}
h and k lines, can add additional information in the lower chromosphere region
without significantly impacting the fitting to the \ion{Mg}{2} h and k lines.
To include the UV triplet lines we add their two upper levels to the Mg
model described in Sec.~\ref{sec3}. The Mn atomic model has nine levels, four
\ion{Mn}{1} levels, four \ion{Mn}{2} levels, and the ground level of \ion{Mn}{3}.

In Figs.~\ref{fig4a}--\ref{fig3a} we show the result of the inversion
(following the strategy described in Sec.~\ref{sec3}) for a pixel in the
plage, for a pixel in the vicinity of the penumbra, and for a pixel
in the penumbra (red, yellow, and green ``$\times$'' symbols in panel (E)
of \fref{fig1}), respectively.
The dashed gray curve in the bottom left panel of the figures show the optical
depth (in $\rm log_{10}(\tau_{500})$) where the optical depth at each wavelength 
is equal to one ($\tau_{\lambda}=1$), which roughly indicates the formation regions
of the \ion{Mg}{2} and \ion{Mn}{1} lines: 
the inner lobes of the circular polarization profiles of the h and k lines
form at around \TAUE{-6.0} (upper chromosphere), while their outer lobes form between
around \TAUE{-4.0} and $-5.0$ (middle chromosphere); the lobes of the circular 
polarization profiles of the \ion{Mn}{1} lines and the \ion{Mg}{2} subordinate 
lines form deeper in the atmosphere (lower chromosphere).
The inferred $B_\parallel$, $T$, and $v_{\rm turb}$ are shown in the right column 
of the figure. In \fref{fig2a}, the two inversions return very similar results,
compatible within
the error bars (see Sec.~\ref{sec3}). The fitting of the Stokes profiles of the
\ion{Mn}{1} lines, and the \ion{Mg}{2} UV triplet lines are not as good as for
the h and k lines. Even though the fit could be improved by adding more weight
them, or by adding more nodes in their region of formation, the approximations in
the modeling of the \ion{Mn}{1} lines would lead to a worse fit of the h and k
lines, and thus, additional inaccuracies in the inferred model.

For the inversion in Figs.~\ref{fig4a} and \ref{fig3a} the differences
between the two approaches are more significant, especially for $v_{\rm turb}$. 
Although we think that the inversion including the \ion{Mn}{1} lines and the
\ion{Mg}{2} UV triplet lines (red curves in Figs.~\ref{fig4a} -- \ref{fig3a}) is more
reliable, since the these lines can add better constraints to the model in
the lower chromosphere and especially the \ion{Mg}{2} lines at 279.88~nm are 
sensitive to the temperature increase at the temperature minimum region 
\citep{Pereira2015ApJ}, the inversion results with only 4 Mg levels are also
acceptable, with both solutions of $B_\parallel$ usually being fully compatible 
within the error bars. Therefore, in order to reduce the already significant 
computing time, we perform the inversion of only the h and k lines at 
all the pixels with the four level Mg model. On average, the inversion still 
takes around 10 CPU hours for each pixel. Besides, we have set an upper
limit of 6~km~s$^{-1}$ to the micro-turbulent velocity at \TAUE{-4.0} to avoid
a compatible large value, because the micro-turbulent velocity is typically found
to be smaller than 6~km~s$^{-1}$ at those heights \citep{SainzDalda2019ApJ}.

\bibliography{clasp21}{}

\bibliographystyle{aasjournal}



\end{document}